\documentstyle[12pt]{article}
\begin{document}
\begin{center}
{\Large\bf Near-threshold \mbox{\boldmath $\eta$}
production in the}\\[2ex]
{\Large\bf \mbox{\boldmath $p\,d \to p\,d\,\eta$} reaction}\\[7ex] 
F.~Hibou$^{\,1}$,
C.~Wilkin$^{\,2}$, 
A.M.~Bergdolt$^{\,1}$, 
G.~Bergdolt$^{\,1}$, 
O.~Bing$^{\,1}$, 
M.~Boivin$^{\,3}$,
A.~Bouchakour$^{\,1}$,
F.~Brochard$^{\,3,4}$,
M.P.~Combes-Comets$^{\,5}$,
P.~Courtat$^{\,5}$,
R.~Gacougnolle$^{\,5}$, 
Y.~Le~Bornec$^{\,5}$,  
A.~Moalem$^{\,6}$,
F.~Plouin$^{\,3,4}$,
F.~Reide$^{\,5}$,
B.~Tatischeff$^{\,5}$, 
N.~Willis$^{\,5}$.\\[5ex]
\end{center}
$^1$ Institut de Recherches Subatomiques, IN2P3-CNRS/Universit\'e
Louis Pasteur,\\
\phantom{$^1$} B.P.~28, F--67037 Strasbourg Cedex~2, France\\[1ex]
$^{2}$ University College London, London WC1E 6BT, United Kingdom\\[1ex]
$^3$ Laboratoire National Saturne, F--91191 Gif-sur-Yvette Cedex,
France\\[1ex]
$^{4}$ LPNHE, Ecole Polytechnique, F-91128 Palaiseau, France\\[1ex]
$^{5}$ Institut de Physique Nucl\'eaire, IN2P3--CNRS/Universit\'e Paris-Sud,\\
\phantom{$^3$} F--91406 Orsay Cedex, France\\[1ex]
$^{6}$ Physics Department, Ben Gurion University, 84105 Beer Sheva,
Israel\\[4ex]

\begin{abstract}
\baselineskip 4ex
The total cross section of the $p\,d \to p\,d\,\eta$ reaction has
been measured at two energies near threshold by detecting the final 
proton and deuteron in a magneti  spectrometer. The values are somewhat
larger than expected on the basis of two simple theoretical estimates.
\\[3ex]
\end{abstract}

\noindent
Corresponding author:\\ 
Colin Wilkin,\\ Department of Physics and Astronomy,\\ 
University College London,\\ 
Gower Street, London WC1E 6BT, UK.\\
E-mail: cw@hep.ucl.ac.uk

\newpage
\baselineskip 5ex
The production of $\eta$ mesons in the $p\,d\to^{\,3}$He$\,\eta$
reaction near threshold is remarkable for both its strength and its
energy dependence~\cite{Berger,Mayer}. The threshold amplitude is of a
similar size to that for pion production, despite the much larger
momentum transfers associated with $\eta$ formation. Although the
angular distribution remains isotropic, suggesting $S$-wave production,
the square of the amplitude falls
by a factor of three over a 5~MeV change in the c.m.\ excess energy $Q$.
This has been taken as indicative of a nearby quasi-bound state of the 
$\eta\,^{3}$He system, arising through strong $\eta$ multiple scatterings
from the three nucleons in the recoiling nucleus~\cite{CW}.
 
In order to transfer such large momenta, Kilian and Nann~\cite{Kilian} 
suggested two-step processes involving intermediate pions. They showed that
the threshold kinematics for $p\,d \to\, ^3$He$\,\eta$ were in a sense
{\it magic}. The momentum of the $\eta$ produced in the reaction
is very similar to that obtained from the
sequential physical processes of $pp\to d\pi^+$ followed by $\pi^+n\to
p\eta$, when there is no relative momentum between the final $p\,d$ pair
and all Fermi momenta are neglected. In such cases the final proton
and deuteron are likely to stick to form the observed $^3$He nucleus.
The classical estimate of the enhancement due to the magic kinematics
is broadly confirmed by quantum mechanical calculations~\cite{FW1},
which reproduce the size of the near-threshold cross section to within
about a factor of two.

The same two-step model should also be capable of explaining events where
the final proton and deuteron emerge freely in the $p\,d\to p\,d\,\eta$
reaction and the aim of the present investigation was to undertake a first
exploration of this cross section near the threshold beam energy of
$T_p=901.2$~MeV. Unfortunately, due to the closure of the laboratory, data
could only be obtained at two beam energies.

The experiment was carried out at the Laboratoire National SATURNE
(LNS), using the large acceptance magnetic spectrometer SPESIII, which was
well adapted for the study of meson production in three-body final states
near threshold through the detection of two charged particles. The
experimental conditions concerning the beam monitoring,
particle detection and identification, were rather similar to those of
previous studies of meson production in the $p\,p\to p\,p\,X$, where the
meson $X$ was identified by the missing mass method~\cite{ppX}. A liquid
deuterium target of 207~mg/cm$^2$ thickness was employed and, in order to
improve the missing mass resolution and the signal-to-background ratio, the
opening of the vertical collimators of SPESIII was reduced to $\pm 40$~mr.

One special feature of the $p\,d\to p\,d\,\eta$ reaction near threshold is
the rather low momentum, around 400-500~MeV/c, of the outgoing protons. This
is to be compared with the standard 600-1400 MeV/c momentum range of the
SPESIII spectrometer. The momentum of the recoiling deuterons is
about 900~MeV/c and, in order to detect both particles simultaneously,
the magnetic field was tuned down to accept momenta from about 360 to
960~MeV/c. Under normal SPESIII working conditions, the values of the
particle momenta were obtained by using well established polynomial relations,
taking the coordinates of the trajectories near the focal surface
as input. The properties of SPESIII were not extensively studied
with reduced fields and, in the present experiment, we used the polynomial
parametrisation with the momenta of the particles scaled according to
the ratio of the actual to the standard mean field, (2.03~Tesla)/(3.07~Tesla).
This procedure essentially assumes that the field was reduced uniformly.
A similar method was applied in the simulations, applying the same
ratio to the momenta when tracking the particles.
Such simulations are important for generating the expected missing mass
peak of the $p\,d\to p\,d\,\eta$ reaction as well as the background
spectrum of the $p\,d\to p\,d\,2\pi$ reaction.

Two-dimensional experimental and simulated scatter plots of the 
emerging proton and deuteron momenta are shown in Fig.~1. Superimposed
upon a fairly uniform background, due the $p\,d\to p\,d\,2\pi$
reaction, there is a darker ellipse inside which the $p\,d\to p\,d\,\eta$
events are confined.
The experimental missing mass spectra of the $p\,d\to p\,d\,X$ reaction
are shown in Figs.~2a1 and 2a2. Clear $\eta$ peaks are observed near the 
upper edges of phase space for the two nominal proton beam energies of
905 and 909~MeV. Simulated background spectra of the 
$p\,d\to p\,d\,2\pi$ reaction are shown in Figs.~2b1 and 2b2 and simulated 
peaks of the $p\,d\to p\,d\,\eta$ reaction are represented by the solid 
lines in  Figs.~2c1 and 2c2. To evaluate the number of 
$p\,d\to p\,d\,\eta$ events, the two simulated spectra were combined
so as to fit the experimental data. After subtracting the simulated 
background from the experimental results, the remaining events (points
with error bars) in Figs.~2c1 and 2c2 show good agreement
with the simulated $p\,d\to p\,d\,\eta$ spectra (solid line). 

Taking the mass of the $\eta$ meson to be 547.30~MeV/c$^2$~\cite{Caso}, 
the best fits were obtained by assuming incident proton energies which were
1-2~MeV lower than the nominal values derived from the Saturne machine
parameters. The fits also suggested adjusting the mean field ratio to a 
value slightly below that of the initial 2.03/3.07 {\it ansatz}. However,
due to the uncertainties in the effective field strength and the particle
tracking, no definitive accurate values of the beam energies could be deduced
from the fitting procedure. But, since the shift indicated here is very
similar to the mean difference $\Delta T=T_{\mbox{\tiny nominal}} -
T_{\mbox{\tiny measured}} =1.1\pm1.0$~MeV obtained at Saturne
from other meson-production reactions near threshold~\cite{ppX,Plouin,Willis},
we adopt this energy correction $\Delta T$, to be subtracted from the
nominal values to obtain the `true' ones. The average values of the excess
energies $Q =\sqrt{(m_p+m_d)^2+2m_dT_p}-m_p-m_d-m_{\eta}$, where the $m_i$
are particle masses, were determined using the corrected proton energies and
taking into account energy losses in the target. These energies were used
in the simulations required to evaluate the SPESIII acceptances, which were
estimated assuming phase-space distributions of final particles. The acceptance
decreases very quickly above threshold through one of the final particles 
falling outside the solid angle of SPESIII. Nevertheless, from
the resulting angular dependence of the acceptance shown in Fig.~3 as a
function of the cosine of the $\eta$ emission angle in the centre-of-mass
system, it can be seen that all the regions of phase space were covered.
The resulting overall acceptances of 24\% and 7\%, evaluated at the 
mid-target energies of $T_p=903.7$ and 907.7~MeV respectively, lead to
the cross sections given in Table~1. The first error on the cross sections 
given in the table includes the statistical error (10\% and 18\% respectively) 
and a 13\% systematic error on the absolute normalisation. The 
$\pm 0.6$~MeV uncertainty in $Q$ gives rise to the additional quoted errors 
through the rapid acceptance variation. However, the latter errors affect
little the comparison with theory since, if both values of $Q$ are increased
by $0.6$~MeV, the experimental points move largely in the directions
given by the theoretical curves.

Estimates have been made of the $p\,d\to p\,d\,\eta$ total cross section near
threshold in the quantum two-step model of Ref.~\cite{Ulla}, though
neglecting all final state interactions. The energy variation of
\begin{equation}
\sigma_T(pd\to pd\eta) = 1.2\,Q^2\ \mbox{\rm nb}
\end{equation}
is compatible with that of our data shown in Fig.~4. The predicted values of
2.7 and 17~nb, at $Q=1.5$ and 3.8~MeV respectively, are only about a factor
of two smaller
than our results and this discrepancy could be due to the neglect of the
strong final state interaction between the proton and deuteron.

There are two possible $S$-wave proton-deuteron final states, 
corresponding to spin $\frac{1}{2}$ and $\frac{3}{2}$. The low energy 
spin-quartet scattering wave functions show little structure
at short distances, whereas the spin-doublet bear some similarity to the
shape of the $p\,d$ distribution at short distances inside the bound 
$^3$He nucleus~\cite{Jaume}.

The connection between the bound and scattering $p\,d$ wave functions
can be exploited to estimate the production of $\eta$ mesons in the
$p\,d\to p\,d\,\eta$ reaction, with the final $p\,d$ system in the
spin-$\frac{1}{2}$ state, in terms of the cross section for
$p\,d\to\,^3\mbox{\rm He}\,\eta$. Provided that the structure of the deuteron
is neglected, the relative
normalisation of the bound and scattering wave functions at short distances
is fixed purely by the proton-deuteron binding energy, 
$\epsilon\approx 5.5$~MeV, in the $^3$He nucleus. If the meson production 
operator is also of short range, the production in two and three-body 
final states should be related through~\cite{FW2}
\begin{equation}
\sigma_T(pd\to pd\eta) = \frac{1}{4}\,\left(\frac{Q}{\epsilon}\right)^{3/2}\,
\left(1+\sqrt{1+Q/\epsilon}\,\right)^{-2}\times
\sigma_T(pd\to\,^3\mbox{\rm He}\,\eta)\:.
\end{equation}
This approach~\cite{FW2} reproduces about $\frac{2}{3}$ of the 
$p\,d\to p\,d\,\pi^0$ total cross section~\cite{Rohdjess} in terms of 
that for $p\,d\to\,^3\mbox{\rm He}\,\pi^0$~\cite{Nikulin} and the residue
could be due to spin-$\frac{3}{2}$ final states.

The very precise near-threshold $p\,d\to\,^3$He$\,\eta$ total cross section
data~\cite{Mayer} may be parametrised as
\begin{equation}
\sigma_T(pd\to\,^3\mbox{\rm He}\,\eta) = \left(\frac{p_{\eta}}{p_p}\right)\,
\frac{22}{(1+1.6p_{\eta})^2+(3.8 p_{\eta})^2}\ \
\mu\mbox{\rm b}\:,
\end{equation}
where the $\eta$ and proton c.m.\ momenta $p_{\eta}$ and $p_p$ are measured
in fm$^{-1}$. This parametrisation is shown in Fig.~4.

The predictions of Eq.~(2) for the $p\,d\to p\,d\,\eta$ total cross sections
are about a factor of three lower than our experimental results shown
in Fig.~4. This may be due to the short-range
assumption for the meson-production operator made in deriving Eq.~(2).
In the two-step model of ref.~\cite{FW1}, the momentum transfer
is provided through having a secondary interaction with an intermediate pion.
Since high Fermi momenta are then not required, this means that the final
$p\,d$ system is not necessarily produced at short distances. However,
at larger distances the scattering wave functions are generally bigger than 
bound state ones, which must die off exponentially. It would therefore
be very desirable to have a microscopic two-step model
calculation of the type of ref.~\cite{Ulla} but with the proton-deuteron
final state interaction included.

We have made the first measurements of the $p\,d\to p\,d\,\eta$ reaction
near threshold and obtained cross sections about a factor of 2-3 higher
than those of two simple theoretical approaches. Since, for our data,
$Q$ is less than the $^3$He binding energy $\epsilon$,
the difference in the energy dependence of the two present models
comes principally from the striking behaviour of the
$pd\to\,^3\mbox{\rm He}\,\eta$ cross section. More detailed
experiments, with a much better resolution in $Q$, are required to see
if there is in fact a strong $\eta$ final state interaction
in the $p\,d\,\eta$ system to match that in $^3\mbox{\rm He}\,\eta$.

We wish to thank the Saturne accelerator crew and support staff for providing
us with working conditions which led to the present results. Discussions
with U.~Tengblad regarding Ref.~\cite{Ulla} were very useful.

\newpage
\begin{table}[hb]
\caption{Measured total cross sections for the $p\,d\to p\,d\,\eta$
reaction at a mid-target centre-of-mass excess energy $Q$. Although
there are $\pm 0.6$~MeV uncertainties in the values of $Q$, the
relative value is correct to $\pm 0.1$~MeV. The first quoted
error in the cross sections includes statistical and normalisation 
uncertainties; the second is that induced by the uncertainty in $Q$. 
For the first point the energy uncertainties are negligible for 
$\sigma/Q$, whereas for the second point it is $\sigma/Q^2$ which is 
largely unaffected by $\Delta Q$.}
\begin{center}
\vspace{2ex}
\begin{tabular}{|c|c|}
\hline
&\\
$Q$&$\sigma_T(pd\to pd\eta)$\\
(MeV)&(nb)\\
&\\
\hline
&\\
1.5$\pm$0.6&6.1$\pm$1.0$\pm^{2.8}_{2.0}$\\
&\\
\hline
&\\
3.8$\pm$0.6&40$\pm$9$\pm^{13}_{10}$\\
&\\
\hline
\end{tabular}
\end{center}
\end{table}
\noindent
\input epsf
\begin{figure}[hb]
\begin{center}
\mbox{\epsfxsize=3.5in \epsfbox{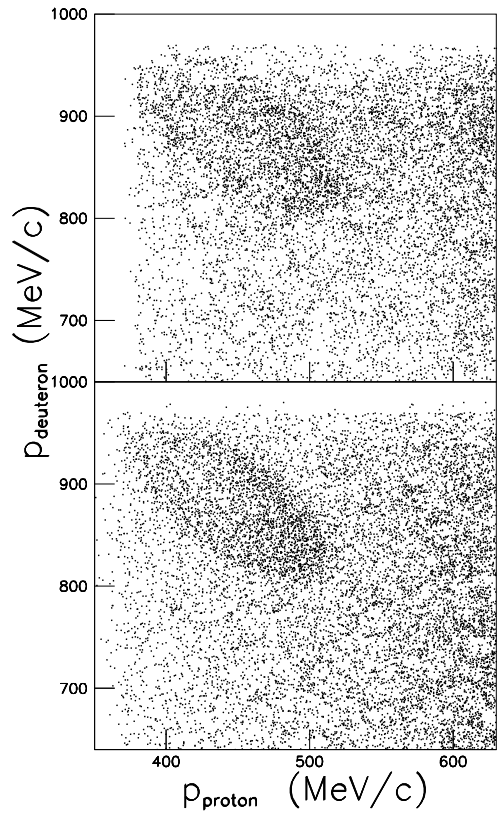}}
\end{center}
\caption{
Two-dimensional scatter plot of the momenta of the proton and deuteron
arising from the reaction $p\,d\to p\,d\,X$ a little above the $\eta$ 
threshold. The experimental data, shown in the upper figure, were
measured at a nominal beam energy $T_p=909$~MeV but, as discussed in
the text, this is expected to be 1-2~MeV above the true value. The
simulation in the lower figure was carried out at 907~MeV. Events from
the $p\,d\to p\,d\,2\pi$ reaction populate the whole plot but in
the top left corner one can see an ellipse containing extra events
corresponding to $\eta$ production.}
\label{fig_1}
\end{figure}
\noindent
\input epsf
\begin{figure}[hb]
\begin{center}
\mbox{\epsfxsize=5.5in \epsfbox{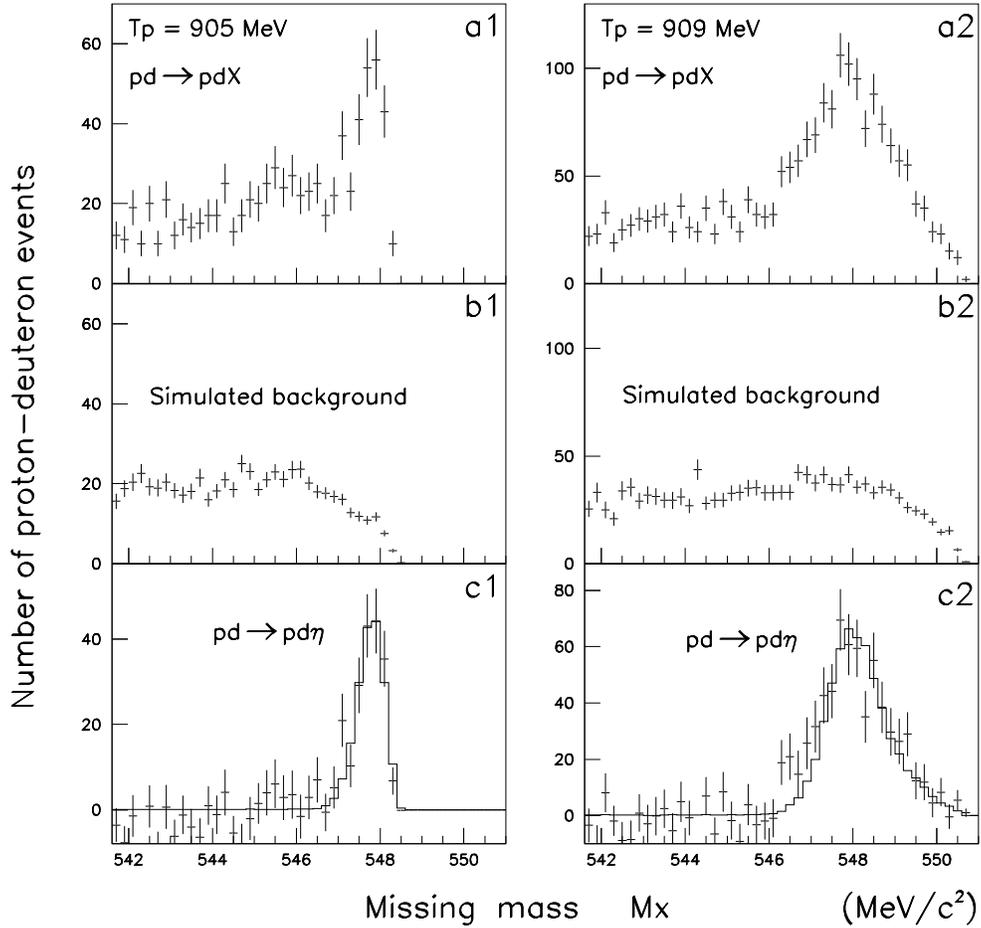}}
\end{center}
\caption{Missing mass spectra of the $p\,d\to p\,d\,X$ reaction at nominal beam
energies of (1) 905~MeV, and (2) 909~MeV. The observed spectra in
(a1) and (a2) are to be compared with background spectra $p\,d\to p\,d\,2\pi$
in (b1) and (b2) estimated by Monte Carlo simulation. The shapes 
and widths of the $\eta$ peaks, found by subtraction and shown in 
(c1) and (c2), are in good agreement with the Monte Carlo predictions
(solid lines) assuming true beam energies 1-2~MeV below the nominal 
values suggested by the macroscopic parameters of Saturne.}
\label{fig_2}
\end{figure}
\noindent
\input epsf
\begin{figure}[hb]
\begin{center}
\mbox{\epsfxsize=5.5in \epsfbox{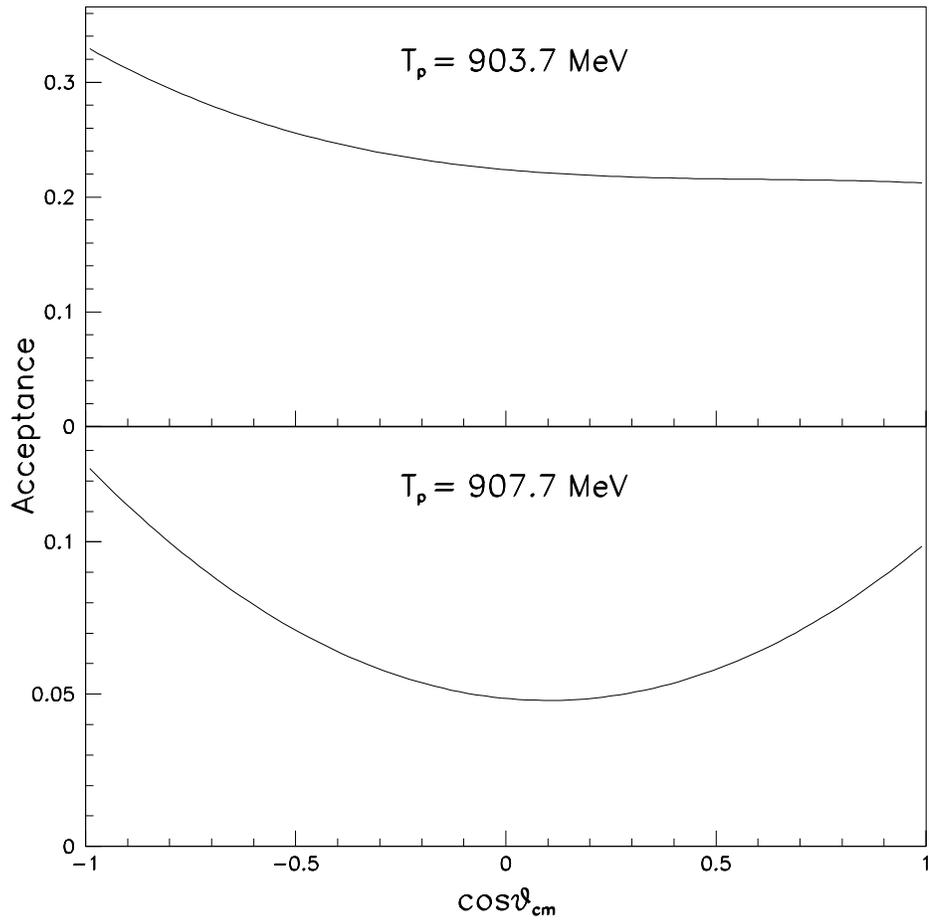}}
\end{center}
\caption{Acceptance of the SPESIII spectrometer for the reaction 
$p\,d\to p\,d\,\eta$ at mid-target beam energies of 903.7 and 907.7~MeV
as a function of the cosine of the $\eta$ production angle in the
centre-of-mass system. The curves have been calculated assuming
phase-space distributions.}
\label{fig_3}
\end{figure}
\noindent
\input epsf
\begin{figure}[hb]
\begin{center}
\mbox{\epsfxsize=5.5in \epsfbox{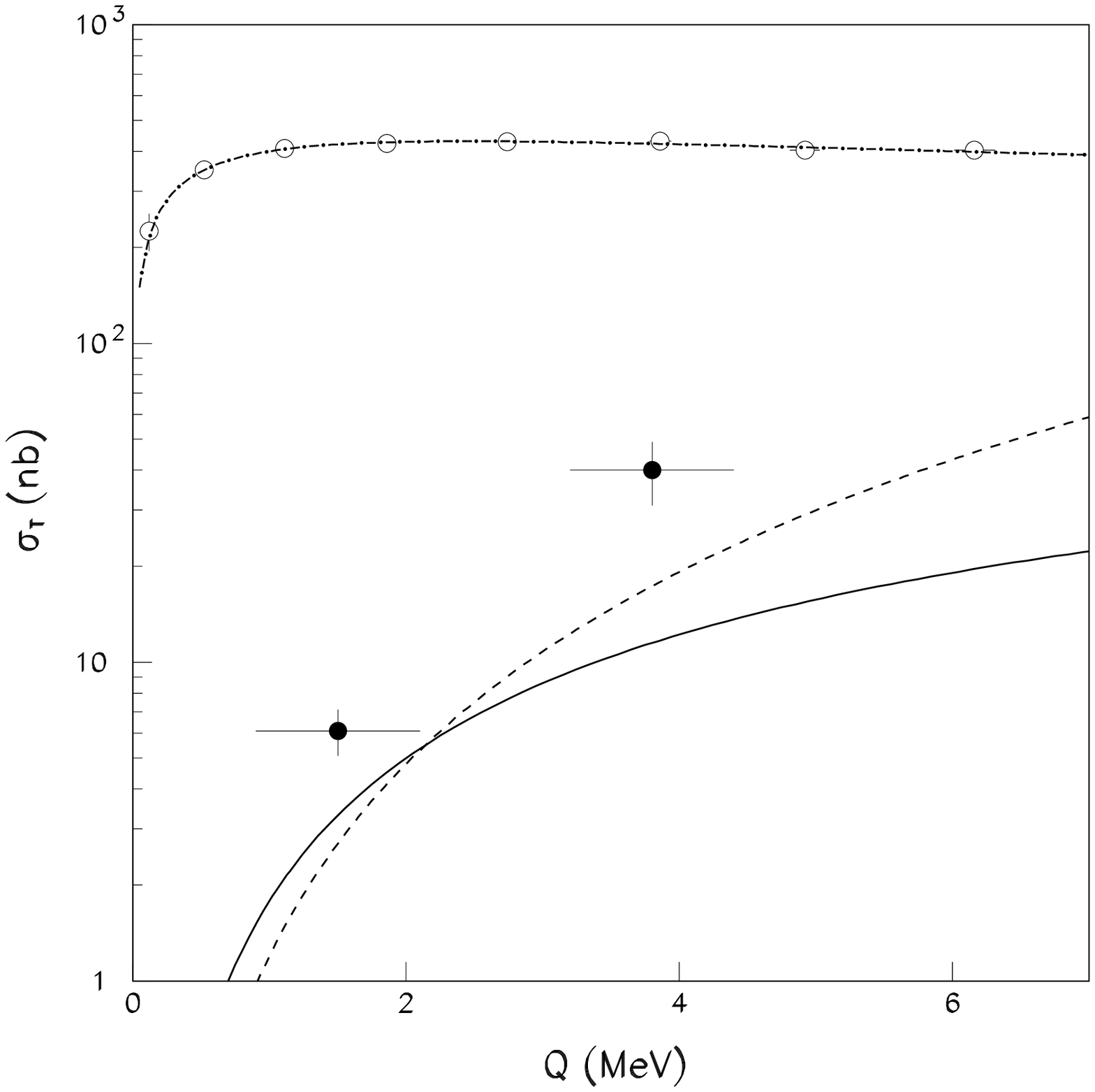}}
\end{center}
\caption{Measured total cross sections for the $p\,d\to p\,d\,\eta$ reaction
close to threshold. The additional errors arising from the influence
of the uncertainty in the absolute value of $Q$ given in Table~1 are not
indicated since, due to the $Q-\sigma$ error correlation, they have much
smaller components perpendicular to the theoretical predictions. The solid
curve is the prediction for the spin-$\frac{1}{2}$ final state in the 
final-state-interaction model of eq.~(2)~\protect\cite{FW2}. The dashed 
curve has been estimated in the two-step model of ref.~\protect\cite{Ulla}, 
where final state interactions have been neglected. The 
$pd\to\,^3\mbox{\rm He}\,\eta$ data of ref.~\protect\cite{Mayer} and
the parametrisation of Eq.~(3) are shown by the open circles and 
dot-dashed curve respectively.}
\label{fig_4}
\end{figure}

\newpage
\begin{thebibliography}{99}
%
\bibitem{Berger} J.~Berger {\it et al.}, Phys.\ Rev.\ Lett.\ {\bf 61}
(1988) 919.
%
\bibitem{Mayer} B.~Mayer {\it et al.}, Phys.\ Rev.\ {\bf C53} (1996) 2068.
%
\bibitem{CW} C.~Wilkin, Phys.\ Rev.\ {\bf C47} (1993) R938.
%
\bibitem{Kilian} K.~Kilian and H.~Nann, AIP Conf.\ Rep.\ {\bf 221} (1990) 185.
%
\bibitem{FW1} G.~F\"aldt and C.~Wilkin, Nucl.\ Phys.\ {\bf A587} (1995) 769.
%
\bibitem{ppX} A.M.~Bergdolt {\it et al.}, Phys.\ Rev.\ D {\bf 48} (1993) R2969;
              A.~Taleb, PhD thesis, Universit\'e Louis Pasteur, Strasbourg
              (1994) (CRN 94-61);
              F.~Hibou {\it et al.}, Phys.\ Lett.\ B {\bf 438} (1998) 41;
              F.~Hibou {\it et al.}, Phys.\ Rev.\ Lett.\ {\bf 83} (1999) 492.
%
\bibitem{Caso} C.~Caso {\it et al.}, Eur.\ Phys.\ J.\ {\bf C3} (1998) 1.
%
\bibitem{Plouin} F.~Plouin {\it et al.}, Phys.\ Lett.\ B {\bf 276} (1992) 526.
%
\bibitem{Willis} N.~Willis {\it et al.}, Phys.\ Lett.\ B {\bf 406} (1997) 14.
%
\bibitem{Ulla} U.~Tengblad, TSL/ISV Report 96-0163, 1996 (unpublished).
%
\bibitem{Jaume} J.~Carbonell and M.~Mangin-Brinet, private
communication (1999).
%
\bibitem{FW2} G.~F\"aldt and C.~Wilkin, Phys.Lett. {\bf B382} (1996) 209.
%
\bibitem{Rohdjess} H.~Rohdjess {\it et al.}, Phys.\ Rev.\ Lett.\
{\bf 70} (1993) 2864.
%
\bibitem{Nikulin} V.N.~Nikulin {\it et al.}, Phys.\ Rev.\ {\bf C54} (1996)
1732.
%
\end{thebibliography}
\end{document}